\documentstyle[prd,aps,preprint,tighten,epsfig]{revtex}

\begin{document}

\draft
\preprint{\begin{tabular}{r}
{\bf hep-ph/0102021} \\
{LMU-01-01} \\
{~}
\end{tabular}}

\title{\Large\bf Sum Rules of Neutrino Masses and CP Violation \\ 
in the  Four-Neutrino Mixing Scheme}
\author{\bf Zhi-zhong Xing}
\address{Sektion Physik, Universit$\it\ddot{a}$t M$\it\ddot{u}$nchen,
Theresienstrasse 37A, 80333 M$\it\ddot{u}$nchen, Germany \\
and \\
Theory Division, Institute of High Energy Physics, P.O. Box 918, 
Beijing 100039, China \\
({\it Electronic address: xing@theorie.physik.uni-muenchen.de}) }
\maketitle

\begin{abstract}
We show that the commutator of lepton mass matrices is invariant under
terrestial matter effects in the four-neutrino mixing scheme.
A set of model-independent sum rules for neutrino masses, which may
be generalized to hold for an arbitrary number of neutrino families,
are for the first time uncovered. Useful sum rules for the 
rephasing-invariant measures of leptonic CP violation have also been
found. Finally we present a generic formula of T-violating asymmetries
and expect it to be applicable to the future long-baseline 
neutrino oscillation experiments.
\end{abstract}

\pacs{PACS number(s): 14.60.Pq, 13.10.+q, 25.30.Pt} 

\newpage

{\large\bf 1} ~
Recently the Super-Kamiokande Collaboration has reported some robust 
evidence for atmospheric and solar neutrino oscillations \cite{SK98}. 
In addition, $\overline{\nu}_\mu \rightarrow \overline{\nu}_e$ and 
$\nu_\mu \rightarrow \nu_e$ oscillations have been observed by
the LSND Collaboration \cite{LSND}. A simultaneous interpretation of
solar, atmospheric and LSND neutrino oscillation data has to invoke the
existence of a light sterile neutrino \cite{4N}, because they involve three 
distinct mass-squared differences 
($\Delta m^2_{\rm sun} \ll \Delta m^2_{\rm atm} \ll \Delta m^2_{\rm LSND}$).
In the four-neutrino mixing scheme, CP violation is
generally expected to manifest itself. 
To measure leptonic CP- and T-violating effects needs a new generation of 
accelerator neutrino experiments with very long baselines, including the
proposed neutrino factories \cite{LB}. In such long-baseline experiments 
the earth-induced matter effects, which are likely to 
deform the neutrino oscillation patterns in vacuum and to fake
the genuine CP- and T-violating signals, must be taken into account.

To pin down the underlying dynamics of lepton mass generation and CP violation
relies crucially upon how accurately the fundamental parameters of lepton
flavor mixing can be measured and disentangled from terrestrial matter 
effects \cite{Xing00}.
It is therefore desirable to explore possible model-independent relations
between the effective neutrino masses in matter and the genuine neutrino masses 
in vacuum. It is also useful to establish some 
model-independent relations between the rephasing-invariant measures of CP 
violation in matter and those in vacuum. So far, however, the work of this nature 
has been lacking.

This paper aims to show that the commutator of lepton mass matrices in the 
four-neutrino mixing scheme is invariant under terrestrial matter effects. 
As a consequence, some concise sum rules of neutrino masses can be
obtained model-independently. Such sum rules can even be generalized to
hold for an arbitrary
number of neutrino families. Another set of sum rules are derived as well
for the rephasing-invariant measures of leptonic CP violation in matter.
Finally we present a generic formula of T-violating asymmetries, which
is applicable in particular to the future long-baseline
neutrino oscillation experiments.

\vspace{0.3cm}

{\large\bf 2} ~ 
The phenomenon of lepton flavor mixing arises from the mismatch between the
diagonalization of the charged lepton mass matrix $M_l$ and that of the
neutrino mass matrix $M_\nu$ in an arbitrary flavor basis. Without loss of
generality, one may choose to identify the flavor eigenstates of charged 
leptons with their mass eigenstates. In this specific basis, where $M_l$
is diagonal, 
the lepton flavor mixing matrix $V$ links the neutrino flavor eigenstates 
directly to the neutrino mass eigenstates. For the admixture of one sterile 
($\nu_s$) and three active ($\nu_e, \nu_\mu, \nu_\tau$) neutrinos, 
the explicit form of $V$ reads
\begin{equation}
\left ( \matrix{
\nu_s \cr
\nu_e \cr
\nu_\mu \cr
\nu_\tau \cr} \right ) \; = \; \left ( \matrix{
V_{s0} & V_{s1} & V_{s2} & V_{s3} \cr
V_{e0} & V_{e1} & V_{e2} & V_{e3} \cr
V_{\mu 0} & V_{\mu 1} & V_{\mu 2} & V_{\mu 3} \cr
V_{\tau 0} & V_{\tau 1} & V_{\tau 2} & V_{\tau 3} \cr} \right ) 
\left ( \matrix{
\nu_0 \cr 
\nu_1 \cr
\nu_2 \cr
\nu_3 \cr} \right ) \; ,
%       (1)
\end{equation}
where $\nu_i$ (for $i=0,1,2,3$) denote the mass eigenstates of four neutrinos.
The effective Hamiltonian responsible for the propagation of neutrinos
in vacuum can be written as \cite{Wolfenstein}
\begin{equation}
{\cal H}_{\rm eff} \; =\; \frac{1}{2E} \left (M_\nu M^\dagger_\nu \right )
\; =\; \frac{1}{2E} \left (V D^2_\nu V^\dagger \right ) \; ,
%	(2)
\end{equation}
where $D_\nu \equiv {\rm Diag}\{m_0, m_1, m_2, m_3 \}$, 
$m_i$ are the neutrino mass eigenvalues, and $E \gg m_i$ denotes
the neutrino beam energy. Note that it is $M_\nu M^\dagger_\nu$
(instead of $M^\dagger_\nu M_\nu$) that associates with 
${\cal H}_{\rm eff}$ \cite{Xing01},
no matter whether neutrinos are Dirac or Majorana particles.
When active neutrinos travel through a normal material medium (e.g., the earth), 
which consists of electrons but of no muons or taus, they encounter
both charged- and neutral-current interactions with electrons. 
The neutral-current interaction is universal for $\nu_e$, $\nu_\mu$ 
and $\nu_\tau$, while the charged-current interaction is
associated only with $\nu_e$. Their effects on the mixing and propagating
features of neutrinos have to be taken into account in
all long-baseline neutrino oscillation experiments. 
Let us use ${\tilde M}_\nu$ and $\tilde V$ to denote 
the effective neutrino mass matrix and the effective flavor mixing matrix 
in matter, respectively. Then the effective Hamiltonian responsible 
for the propagation of neutrinos in matter can be written as 
\begin{equation}
\tilde{\cal H}_{\rm eff} \; =\; \frac{1}{2E}
\left (\tilde{M}_\nu \tilde{M}^\dagger_\nu \right ) \; =\; \frac{1}{2E}
\left (\tilde{V} \tilde{D}^2_\nu \tilde{V}^\dagger \right ) \; ,
%	(3)
\end{equation}
where $\tilde{D}_\nu \equiv {\rm Diag}
\{\tilde{m}_0, \tilde{m}_1, \tilde{m}_2, \tilde{m}_3 \}$, and
$\tilde{m}_i$ are the effective neutrino mass eigenvalues in matter.
The deviation of $\tilde{\cal H}_{\rm eff}$ from ${\cal H}_{\rm eff}$
is given by
\begin{equation}
\Delta {\cal H}_{\rm eff} \; \equiv \; \tilde{\cal H}_{\rm eff}
- {\cal H}_{\rm eff}  =
\left ( \matrix{
a' & 0 & 0 & 0 \cr
0 & a & 0 & 0 \cr
0 & 0 & 0 & 0 \cr
0 & 0 & 0 & 0 \cr} \right ) \; ,
%	(4)
\end{equation}
where $a = \sqrt{2} ~ G_{\rm F} N_e$ and
$a' = \sqrt{2} ~ G_{\rm F} N_n/2$ with 
$N_e$ and $N_n$ being the background densities of electrons and 
neutrons \cite{Grimus}, respectively. In the following we assume a
constant earth density profile (i.e., $N_e$ = constant and
$N_n$ = constant), which is a very good approximation for all of the 
presently-proposed long-baseline neutrino experiments.

Now let us introduce the commutators of $4\times 4$ lepton mass matrices to
describe the flavor mixing of one sterile and three active neutrinos. 
Without loss of any generality, we continue to work in the afore-chosen 
flavor basis, where $M_l$ takes the diagonal form
$D_l = {\rm Diag} \{ m_s, m_e, m_\mu, m_\tau \}$ 
with $m_s = 0$. Note that we have assumed the $(1, 1)$ element of 
$D_l$ to be zero, because there is no counterpart of the sterile 
neutrino $\nu_s$ in the charged lepton sector. We shall see later on that
our physical results are completely independent of $m_s$, no matter what 
value it may take. The commutator of lepton mass matrices in vacuum and 
that in matter can then be defined as
\begin{eqnarray}
C & \equiv & i \left [ M_\nu M^\dagger_\nu ~ , M_l M^\dagger_l \right ]
= i \left [V D^2_\nu V^\dagger , D^2_l \right ] \; , 
\nonumber \\
\tilde{C} & \equiv & i \left [ \tilde{M}_\nu \tilde{M}^\dagger_\nu ~ , 
M_l M^\dagger_l \right ] =
i \left [ \tilde{V} \tilde{D}^2_\nu \tilde{V}^\dagger , 
D^2_l \right ] \; .
%	(5)
\end{eqnarray}
Obviously $C$ and $\tilde C$ are traceless Hermitian matrices. In terms
of neutrino masses and flavor mixing matrix elements, we obtain the
explicit expressions of $C$ and $\tilde C$ as follows:
\begin{eqnarray}
C & = & i \left ( \matrix{
0 & \Delta_{es} Z_{se} & \Delta_{\mu s} Z_{s \mu} & \Delta_{\tau s} Z_{s \tau} \cr
\Delta_{se} Z_{es} & 0 & \Delta_{\mu e} Z_{e \mu} & \Delta_{\tau e} Z_{e \tau} \cr
\Delta_{s \mu} Z_{\mu s} & \Delta_{e \mu} Z_{\mu e} & 0 
& \Delta_{\tau \mu} Z_{\mu \tau} \cr
\Delta_{s \tau} Z_{\tau s} & \Delta_{e \tau} Z_{\tau e} 
& \Delta_{\mu \tau} Z_{\tau \mu} & 0 \cr} \right ) \; ,
\nonumber \\ \nonumber \\
\tilde{C} & = & i \left ( \matrix{
0 & \Delta_{es} \tilde{Z}_{se} & \Delta_{\mu s} \tilde{Z}_{s \mu} 
& \Delta_{\tau s} \tilde{Z}_{s \tau} \cr
\Delta_{se} \tilde{Z}_{es} & 0 & \Delta_{\mu e} \tilde{Z}_{e \mu} 
& \Delta_{\tau e} \tilde{Z}_{e \tau} \cr
\Delta_{s \mu} \tilde{Z}_{\mu s} & \Delta_{e \mu} \tilde{Z}_{\mu e} & 0 
& \Delta_{\tau \mu} \tilde{Z}_{\mu \tau} \cr
\Delta_{s \tau} \tilde{Z}_{\tau s} & \Delta_{e \tau} \tilde{Z}_{\tau e} 
& \Delta_{\mu \tau} \tilde{Z}_{\tau \mu} & 0 \cr} \right ) \; ,
%	(6)
\end{eqnarray}
where $\Delta_{\alpha \beta} \equiv m^2_\alpha - m^2_\beta$ for
$\alpha \neq \beta$ running over $(s, e, \mu, \tau)$, and
\begin{eqnarray}
Z_{\alpha \beta} & \equiv & \sum^3_{i=0} \left ( m^2_i V_{\alpha i} 
V^*_{\beta i} \right ) \; ,
\nonumber \\
\tilde{Z}_{\alpha \beta} & \equiv & \sum^3_{i=0} \left ( \tilde{m}^2_i
\tilde{V}_{\alpha i} \tilde{V}^*_{\beta i} \right ) \; .
%	(7)
\end{eqnarray}
One can see that $\Delta_{\beta \alpha} = - \Delta_{\alpha \beta}$,
$Z_{\beta \alpha} = Z^*_{\alpha \beta}$ and
$\tilde{Z}_{\beta \alpha} = \tilde{Z}^*_{\alpha \beta}$ hold.

To find out how $\tilde{Z}_{\alpha \beta}$ is connected with $Z_{\alpha \beta}$,
we need to establish the relation between $\tilde C$ and $C$. 
Taking Eqs. (2), (3) and (4) into account, we immediately obtain
\begin{equation}
\tilde{C} \; = \; 2iE
\left [ \tilde{\cal H}_{\rm eff} ~ , D^2_l \right ] 
= C + 2iE \left [ \Delta {\cal H}_{\rm eff} ~ , D^2_l \right ]  
= C \; .
%	(8)
\end{equation}
This interesting result indicates that {\it the commutator
of lepton mass matrices in vacuum is invariant under terrestrial matter effects}.
As a straightforward consequence of $\tilde{C} = C$, we arrive at
$\tilde{Z}_{\alpha \beta} = Z_{\alpha \beta}$ from Eq. (6). Then a set of
concise sum rules of neutrino masses emerge:
\begin{equation}
\sum^3_{i=0} \left ( \tilde{m}^2_i \tilde{V}_{\alpha i} 
\tilde{V}^*_{\beta i} \right ) \; = \; 
\sum^3_{i=0} \left ( m^2_i V_{\alpha i} V^*_{\beta i} \right ) \; ;
%	(9)
\end{equation}
or equivalently 
\begin{eqnarray}
~~~ \sum^3_{i=1} \left (\tilde{\Delta}_{i0}
\tilde{V}_{\alpha i} \tilde{V}^*_{\beta i} \right ) \; = \;
\sum^3_{i=1} \left ( \Delta_{i0} V_{\alpha i} V^*_{\beta i} \right ) \; ,
%	(10)
\end{eqnarray}
where $\Delta_{i0} \equiv m^2_i - m^2_0$ and 
$\tilde{\Delta}_{i0} \equiv \tilde{m}^2_i - \tilde{m}^2_0$ for $i=1,2,3$.
It becomes obvious that the validity of Eq. (9) or Eq. (10) has nothing to
do with the assumption of $m_s = 0$ in the charged lepton sector.
Although we have derived these sum rules in the four-neutrino mixing scheme,  
they may simply be generalized to hold for an arbitrary number of neutrino families.

It should be noted that $Z_{\alpha \beta}$ and $\tilde{Z}_{\alpha \beta}$
are sensitive to a redefinition of the phases of charged lepton fields. The
simplest rephasing-invariant equality is of course  
$|\tilde{Z}_{\alpha \beta}| = |Z_{\alpha \beta}|$. For the description of
CP or T violation in neutrino oscillations, we are more interested in the following 
rephasing-invariant relationship:
\begin{equation}
\tilde{Z}_{\alpha \beta} \tilde{Z}_{\beta \gamma} \tilde{Z}_{\gamma \alpha}
\; =\; Z_{\alpha \beta} Z_{\beta \gamma} Z_{\gamma \alpha} \; ,
%	(11)
\end{equation}
for $\alpha \neq \beta \neq \gamma$ running over $(s, e, \mu, \tau)$.
As one can see later on, 
the imaginary parts of $Z_{\alpha \beta} Z_{\beta \gamma} Z_{\gamma \alpha}$
and $\tilde{Z}_{\alpha \beta} \tilde{Z}_{\beta \gamma} \tilde{Z}_{\gamma \alpha}$
are related respectively to leptonic CP violation in vacuum and that in matter.

It should also be noted that the results obtained above are only valid for
neutrinos propagating in vacuum and in matter. As for antineutrinos, the
corresponding formulas can straightforwardly be written out from 
Eqs. (3) -- (11) through the replacements $V\Longrightarrow V^*$, 
$a \Longrightarrow -a$ and $a' \Longrightarrow -a'$.

\vspace{0.3cm}

{\large\bf 3} ~
In terms of the matrix elements of $V$ or $\tilde V$, one may define the
rephasing-invariant measures of CP violation as follows \cite{Dai}:
\begin{eqnarray}
J^{ij}_{\alpha\beta} & \equiv & {\rm Im} \left ( V_{\alpha i} V_{\beta j}
V^*_{\alpha j} V^*_{\beta i} \right ) \; , 
\nonumber \\
\tilde{J}^{ij}_{\alpha\beta} & \equiv & {\rm Im} \left ( \tilde{V}_{\alpha i} 
\tilde{V}_{\beta j} \tilde{V}^*_{\alpha j} \tilde{V}^*_{\beta i} \right ) \; , 
%	(12)
\end{eqnarray}
where the Greek subscripts ($\alpha \neq \beta$) run over $(s, e, \mu, \tau)$,
and the Latin superscripts ($i\neq j$) run over $(0, 1, 2, 3)$. Of course,
$J^{ii}_{\alpha\beta} = J^{ij}_{\alpha\alpha} =0$ and
$\tilde{J}^{ii}_{\alpha\beta} = \tilde{J}^{ij}_{\alpha\alpha} =0$ hold by
definition. With the help of the unitarity of $V$ or $\tilde V$, one may 
straightforwardly obtain the following correlation equations of 
$J^{ij}_{\alpha \beta}$ and $\tilde{J}^{ij}_{\alpha\beta}$:
\begin{eqnarray}
\sum_i J^{ij}_{\alpha\beta} & = & 
\sum_j J^{ij}_{\alpha\beta} \;\; =\;\; 0 \;\; ,
\nonumber \\
\sum_i \tilde{J}^{ij}_{\alpha\beta} & = & 
\sum_j \tilde{J}^{ij}_{\alpha\beta} \;\; =\;\; 0 \;\; ;
%	(13)
\end{eqnarray}
and
\begin{eqnarray}
\sum_\alpha J^{ij}_{\alpha\beta} & = & 
\sum_\beta J^{ij}_{\alpha\beta} \;\; =\;\; 0 \;\; ,
\nonumber \\
\sum_\alpha \tilde{J}^{ij}_{\alpha\beta} & = & 
\sum_\beta \tilde{J}^{ij}_{\alpha\beta} \;\; =\;\; 0 \;\; .
%	(14)
\end{eqnarray}
Note that there are totally nine independent 
$J^{ij}_{\alpha\beta}$ (or $\tilde{J}^{ij}_{\alpha\beta}$) in the 
four-neutrino mixing scheme under discussion. If only the flavor 
mixing of three active neutrinos is taken into account, there will
be a single independent $J^{ij}_{\alpha\beta}$ 
(or $\tilde{J}^{ij}_{\alpha\beta}$):
\begin{eqnarray}
J^{ij}_{\alpha\beta} ~ & = & ~ J \sum_k
\epsilon_{ijk} \sum_\gamma \epsilon_{\alpha\beta\gamma} \; ,
\nonumber \\
\tilde{J}^{ij}_{\alpha\beta} ~ & = & ~ \tilde{J} \sum_k
\epsilon_{ijk} \sum_\gamma \epsilon_{\alpha\beta\gamma} \; ,
%	(15)
\end{eqnarray}
where $(i,j,k)$ and $(\alpha, \beta, \gamma)$ run over 
$(1,2,3)$ and $(e, \mu, \tau)$, respectively.
It is a unique feature of the three-family flavor mixing scenario, 
for either leptons or quarks \cite{Jarlskog}, 
that there exists a universal CP-violating parameter.

To establish the relationship between $\tilde{J}^{ij}_{\alpha\beta}$ 
and $J^{ij}_{\alpha\beta}$, we make use of th equality in Eq. (11).
The key point is that the imaginary parts of the rephasing-invariant quantities 
$Z_{\alpha\beta} Z_{\beta\gamma} Z_{\gamma\alpha}$ and 
$\tilde{Z}_{\alpha\beta} \tilde{Z}_{\beta\gamma} \tilde{Z}_{\gamma\alpha}$, 
\begin{eqnarray}
{\rm Im} ( Z_{\alpha\beta} Z_{\beta\gamma} Z_{\gamma\alpha} )
& = & \sum^3_{i=1} \sum^3_{j=1} \sum^3_{k=1} \left [ \Delta_{i0} \Delta_{j0}
\Delta_{k0} ~ {\rm Im} \left ( V_{\alpha i} V_{\beta j} V_{\gamma k} 
V^*_{\alpha k} V^*_{\beta i} V^*_{\gamma j} \right ) \right ] \; ,
\nonumber \\
{\rm Im} ( \tilde{Z}_{\alpha\beta} \tilde{Z}_{\beta\gamma} 
\tilde{Z}_{\gamma\alpha} )
& = & \sum^3_{i=1} \sum^3_{j=1} \sum^3_{k=1} \left [ \tilde{\Delta}_{i0} 
\tilde{\Delta}_{j0} \tilde{\Delta}_{k0} ~ {\rm Im}
\left ( \tilde{V}_{\alpha i} \tilde{V}_{\beta j} \tilde{V}_{\gamma k} 
\tilde{V}^*_{\alpha k} \tilde{V}^*_{\beta i} 
\tilde{V}^*_{\gamma j} \right ) \right ] \; ,
%	(16)
\end{eqnarray}
which do not vanish unless leptonic CP and T are good symmetries, 
amount to each other. The right-hand side of Eq. (16) can be expanded
in terms of $J^{ij}_{\alpha\beta}$ and $\tilde{J}^{ij}_{\alpha\beta}$. In
doing so, one needs to use Eqs. (12), (13) and (14) as well as
the unitarity conditions of $V$ and $\tilde{V}$
frequently. After some lengthy but straightforward algebraic calculations,
we arrive at the following sum rules of CP- or T-violating parameters:
\begin{eqnarray}
& & \tilde{\Delta}_{10} \tilde{\Delta}_{20} \tilde{\Delta}_{30} \sum^3_{i=1}
\left ( \tilde{J}^{0i}_{\alpha\beta} |\tilde{V}_{\gamma i}|^2 + 
\tilde{J}^{0i}_{\beta\gamma} |\tilde{V}_{\alpha i}|^2 + 
\tilde{J}^{0i}_{\gamma\alpha} |\tilde{V}_{\beta i}|^2 
\right ) 
\nonumber \\
& & + \sum^3_{i=1} \sum^3_{j=1} \left [ \tilde{\Delta}_{i0} 
\tilde{\Delta}^2_{j0} \left (
\tilde{J}^{ij}_{\alpha\beta} |\tilde{V}_{\gamma j}|^2 + 
\tilde{J}^{ij}_{\beta\gamma} |\tilde{V}_{\alpha j}|^2 + 
\tilde{J}^{ij}_{\gamma\alpha} |\tilde{V}_{\beta j}|^2 
\right ) \right ] 
\nonumber \\
& = & \Delta_{10} \Delta_{20} \Delta_{30} \sum^3_{i=1}
\left ( J^{0i}_{\alpha\beta} |V_{\gamma i}|^2 + 
J^{0i}_{\beta\gamma} |V_{\alpha i}|^2 + J^{0i}_{\gamma\alpha} |V_{\beta i}|^2 
\right ) 
\nonumber \\
& & + \sum^3_{i=1} \sum^3_{j=1} \left [ \Delta_{i0} \Delta^2_{j0} \left (
J^{ij}_{\alpha\beta} |V_{\gamma j}|^2 + 
J^{ij}_{\beta\gamma} |V_{\alpha j}|^2 + J^{ij}_{\gamma\alpha} |V_{\beta j}|^2 
\right ) \right ] \; .
%	(17)
\end{eqnarray}
We remark that this new result is model-independent and 
rephasing-invariant. It may be considerably simplified, once the
hierarchy of neutrino masses and that of flavor mixing angles are 
theoretically assumed or experimentally measured. If one 
``switches off'' the mass of the sterile neutrino and its mixing 
with active neutrinos (i.e., $a' = 0$,
$\Delta_{i0} = m^2_i$, $\tilde{\Delta}_{i0} = \tilde{m}^2_i$,
$J^{0i}_{\alpha\beta} =0$, and $\tilde{J}^{0i}_{\alpha\beta} =0$),
then Eq. (17) turns out to take the form
\begin{equation}
\tilde{J} \tilde{\Delta}_{21} \tilde{\Delta}_{31} \tilde{\Delta}_{32}
\; =\; J \Delta_{21} \Delta_{31} \Delta_{32} \; .
%	(18)
\end{equation}
This elegant relationship has been derived in Refs. \cite{Xing01,Scott} 
with the help of the equality ${\rm Det} (\tilde{C}) = {\rm Det} (C)$, 
instead of Eq. (11), in the three-neutrino mixing scheme. 

The matter-corrected CP-violating parameters $\tilde{J}^{ij}_{\alpha\beta}$
can, at least in principle, be determined from the measurement of 
CP- and T-violating effects in a variety of long-baseline neutrino oscillation 
experiments. The conversion probability of a neutrino $\nu_\alpha$ to another
neutrino $\nu^{~}_\beta$ is given in matter as
\begin{equation}
\tilde{P}(\nu_\alpha \rightarrow \nu^{~}_\beta) \; = \;
\delta_{\alpha\beta} - 4 \sum_{i<j} \left [ {\rm Re} \left (
\tilde{V}_{\alpha i} \tilde{V}_{\beta j} \tilde{V}^*_{\alpha j} 
\tilde{V}^*_{\beta i} \right ) 
\sin^2 \tilde{F}_{ji} \right ] - 2 \sum_{i<j} \left [ \tilde{J}^{ij}_{\alpha\beta}
\sin (2 \tilde{F}_{ji}) \right ] \; ,
%	(19)
\end{equation}
where $\tilde{F}_{ji} \equiv 1.27 \tilde{\Delta}_{ji} L/E$ with
$\tilde{\Delta}_{ji} \equiv \tilde{m}^2_j - \tilde{m}^2_i$, 
$L$ stands for the baseline length (in unit of km), and $E$ is the neutrino
beam energy (in unit of GeV).
The transition probability $\tilde{P}(\nu^{~}_\beta \rightarrow \nu_\alpha)$ 
can directly be read off from Eq. (19), if the replacements 
$\tilde{J}^{ij}_{\alpha\beta} \Longrightarrow -\tilde{J}^{ij}_{\alpha\beta}$ 
are made
%%%%%%%%%%%%%%%%%%%%%%%
\footnote{Note that the differences of effective neutrino masses 
$\tilde{\Delta}_{i0}$ (for $i=1,2,3$), which must be CP-conserving,
keep unchanged for the replacements 
$J^{ij}_{\alpha\beta} \Longrightarrow -J^{ij}_{\alpha\beta}$. Therefore the
sign flip of $J^{ij}_{\alpha\beta}$ results in that of
$\tilde{J}^{ij}_{\alpha\beta}$, as indicated by the sum rules given in Eq. (17).}.
%%%%%%%%%%%%%%%%%%%%%%% 
To obtain the probability
$\tilde{P}(\overline{\nu}_\alpha \rightarrow \overline{\nu}^{~}_\beta)$,
however, both the replacements 
$J^{ij}_{\alpha\beta} \Longrightarrow -J^{ij}_{\alpha\beta}$
and $(a, a') \Longrightarrow (-a, -a')$ need be made for Eq. (19). 
In this case, 
$\tilde{P}(\overline{\nu}_\alpha \rightarrow \overline{\nu}^{~}_\beta)$
is not equal to $\tilde{P}(\nu^{~}_\beta \rightarrow \nu_\alpha)$. The
difference between 
$\tilde{P}(\overline{\nu}_\alpha \rightarrow \overline{\nu}_\beta)$ and
$\tilde{P}(\nu_\beta \rightarrow \nu_\alpha)$ is a false signal of CPT violation,
induced actually by the matter effect \cite{Xing00}. 
Thus the CP-violating asymmetries between
$\tilde{P}(\nu_\alpha \rightarrow \nu^{~}_\beta)$ and
$\tilde{P}(\overline{\nu}_\alpha \rightarrow \overline{\nu}^{~}_\beta)$
are in general different from the T-violating asymmetries between
$\tilde{P}(\nu_\alpha \rightarrow \nu^{~}_\beta)$ and
$\tilde{P}(\nu^{~}_\beta \rightarrow \nu_\alpha)$. The latter
can be explicitly expressed as follows:
\begin{eqnarray}
\Delta \tilde{P}_{\alpha\beta} ~ & \equiv & ~
\tilde{P}(\nu^{~}_\beta \rightarrow \nu^{~}_\alpha) ~ - ~ 
\tilde{P}(\nu^{~}_\alpha \rightarrow \nu^{~}_\beta) \;
\nonumber \\
& = & ~ 4 \left [ \tilde{J}^{01}_{\alpha\beta} \sin (2 \tilde{F}_{10}) +
\tilde{J}^{02}_{\alpha\beta} \sin (2 \tilde{F}_{20}) +
\tilde{J}^{03}_{\alpha\beta} \sin (2 \tilde{F}_{30}) \right . 
\nonumber \\
& & \left . + \tilde{J}^{12}_{\alpha\beta} \sin (2 \tilde{F}_{21}) +
\tilde{J}^{13}_{\alpha\beta} \sin (2 \tilde{F}_{31}) +
\tilde{J}^{23}_{\alpha\beta} \sin (2 \tilde{F}_{32}) \right ] \; .
%	(20) 
\end{eqnarray}
If the hierarchical patterns of neutrino masses and flavor mixing
angles are assumed, the expression of $\Delta \tilde{P}_{\alpha\beta}$ 
may somehow be simplified \cite{Tanimoto}. Note that only three of 
the twelve nonvanishing asymmetries $\Delta \tilde{P}_{\alpha\beta}$
are independent, as a consequence of the unitarity of $\tilde V$ or 
the correlation of $\tilde{J}^{ij}_{\alpha\beta}$. Since
only the transition probabilities of active neutrinos can be 
realistically measured, we are more interested in the T-violating
asymmetries $\Delta \tilde{P}_{e \mu}$, $\Delta \tilde{P}_{\mu \tau}$ and
$\Delta \tilde{P}_{\tau e}$. These three measurables, which are independent of 
one another in the four-neutrino mixing scheme under discussion, must be 
identical in the conventional three-neutrino mixing scheme. 
In the latter case, where $a' = 0$,
$\tilde{J}^{01}_{\alpha\beta} = \tilde{J}^{02}_{\alpha\beta}
= \tilde{J}^{03}_{\alpha\beta} = 0$, and $\tilde{J}^{12}_{\alpha\beta} 
= - \tilde{J}^{13}_{\alpha\beta} = \tilde{J}^{23}_{\alpha\beta} = \tilde{J}$
for $(\alpha, \beta)$ running over $(e, \mu)$, $(\mu, \tau)$ and
$(\tau, e)$, Eq. (20) can be simplified to 
\begin{equation}
\Delta \tilde{P}^{(3)}_{\alpha\beta} \; = \; 16 \tilde{J} 
\sin \tilde{F}_{21} \sin \tilde{F}_{31} \sin \tilde{F}_{32} \; .
%	(21)
\end{equation}
The overall matter contamination residing in $\Delta \tilde{P}_{\alpha\beta}$ 
is usually expected to be insignificant. The reason is simply that the 
terrestrial matter effects in $\tilde{P}(\nu_\alpha \rightarrow \nu^{~}_\beta)$
and $\tilde{P}(\nu^{~}_\beta \rightarrow \nu_\alpha)$, which both depend on the 
parameters $(a, a')$, may partly (even essentially) cancel each 
other in the T-violating asymmetry $\Delta \tilde{P}_{\alpha\beta}$. In contrast, 
$\tilde{P}(\nu_\alpha \rightarrow \nu^{~}_\beta)$ and
$\tilde{P}(\overline{\nu}_\alpha \rightarrow \overline{\nu}^{~}_\beta)$
are associated respectively with $(+a, +a')$ and $(-a, -a')$, thus
there should not have large cancellation of matter effects in the corresponding
CP-violating asymmetries.

\vspace{0.3cm}

{\large\bf 4} ~
In summary, we have introduced the commutators of lepton mass matrices to describe 
the flavor mixing phenomenon of three active and one sterile neutrinos. 
It has been shown that the commutator defined in
vacuum is invariant under terrestrial matter effects. An important
consequence of this interesting result is the emergence of a set of
model-independent sum rules for neutrino masses in two different media.
Such sum rules may even be generalized to hold for an arbitrary number
of neutrino families. We have also uncovered some useful sum rules for the 
rephasing-invariant measures of leptonic CP violation in the
four-neutrino mixing scheme. A generic formula of T-violating asymmetries,
which is applicable in particular to the future long-baseline neutrino
oscillation experiments, has been derived and discussed.

The remarkable strategy of this paper is to formulate the sum rules of
neutrino masses and CP violation in a model-independent way. Hence we 
have concentrated upon the generic formalism instead of the specific 
scenarios. The relevant results are anticipated to be a useful addition 
to the phenomenology of lepton flavor mixing and neutrino oscillations,
and the application of them to the realistic long-baseline neutrino 
experiments deserves a further study elsewhere.


\newpage

\begin{thebibliography}{99}
\bibitem{SK98} Super-Kamiokande Collaboration, 
Y. Fukuda {\it et al.}, Phys. Rev. Lett. {\bf 81}, 1562 (1998); 
{\bf 81}, 4279 (1998); http://www-sk.icrr.u-tokyo.ac.jp/dpc/sk/.

\bibitem{LSND} LSND Collaboration, C. Athanassopoulos {\it et al.},
Phys. Rev. Lett. {\bf 81}, 1774 (1998); Phys. Rev. C {\bf 58}, 2489 (1998).

\bibitem{4N} See, e.g., 
D. Caldwell and R.N. Mohapatra, Phys. Rev. D {\bf 48}, 3259 (1993);
V. Barger, S. Pakvasa, T.J. Weiler, and
K. Whisnant, Phys. Rev. D {\bf 58}, 093016 (1998);
S. Gibbons, R.N. Mohapatra, S. Nandi, and A. Raichoudhury,
Phys. Lett. B {\bf 430}, 296 (1998);
S.M. Bilenky, C. Giunti, W. Grimus, and T. Schwetz,
Phys. Rev. D {\bf 60}, 073007 (1999);
V. Barger, B. Kayser, J. Learned, T. Weiler, and K. Whisnant, 
Phys. Lett. B {\bf 489}, 345 (2000);
C. Giunti and M. Laveder, hep-ph/0010009;
O.L.G. Peres and A.Yu. Smirnov, hep-ph/0011054.

\bibitem{LB} See, e.g., B. Autin {\it et al.}, Report No. 
CERN-SPSC-98-30 (1998); 
M.G. Catanesi {\it et al.}, Report No. CERN-SPSC-99-35 (1999);
D. Ayres {\it et al.}, physics/9911009; C. Albright {\it et al.}, 
hep-ex/0008064; and references therein.

\bibitem{Xing00} Z.Z. Xing, Phys. Lett. B {\bf 487}, 327 (2000).

\bibitem{Wolfenstein} L. Wolfenstein, Phys. Rev. D {\bf 17}, 2369 (1978);
S.P. Mikheyev and A.Yu. Smirnov, Yad. Fiz. (Sov. J. Nucl. Phys.) {\bf 42}, 
1441 (1985).

\bibitem{Xing01} Z.Z. Xing, Phys. Rev. D. {\bf 63}, 073012 (2001).

\bibitem{Grimus} S.M. Bilenky, C. Giunti, and W. Grimus,
Prog. Part. Nucl. Phys. {\bf 43}, 1 (1999);
D. Dooling, C. Giunti, K. Kang, and C.W. Kim, Phys. Rev. D {\bf 61}, 073011 (2000).

\bibitem{Dai} V. Barger, Y.B. Dai, K. Whisnant, and B.L. Young,
Phys. Rev. D {\bf 59}, 113010 (1999).

\bibitem{Jarlskog} C. Jarlskog, Phys. Rev. Lett. {\bf 55}, 1039 (1985);
H. Fritzsch and Z.Z. Xing, Nucl. Phys. B {\bf 556}, 49 (1999);
Prog. Part. Nucl. Phys. {\bf 45}, 1 (2000); hep-ph/9912358.

\bibitem{Scott} P.F. Harrison and W.G. Scott, Phys. Lett. B {\bf 476}, 349 (2000); 
P.I. Krastev and S.T. Petcov, Phys. Lett. B {\bf 205}, 84 (1988).

\bibitem{Tanimoto} A. Kalliom$\rm\ddot{a}$ki, J. Maalampi, and
M. Tanimoto, Phys. Lett. B {\bf 469}, 179 (1999).

\end{thebibliography}
\end{document}